\documentclass[prb,onecolumn,a4paper,amsmath,amssymb,%
superscriptaddress,floatfix,11pt]{revtex4}

\usepackage{amsmath,amssymb}
\usepackage{graphicx,epstopdf}
\newcommand{\Op}[1]{\boldsymbol{\mathsf{\hat{#1}}}}

\newcommand{\Fkt}[1]{\,\mathsf {#1}}
\newcommand{\half}{\frac{1}{2}}

\makeatletter
\renewcommand\section{\@startsection
  {section}
  {1}
  {\z@}
  {0.8cm \@plus1ex \@minus .2ex}
  {0.5cm}
  {\normalfont\bfseries}
}
\makeatother
\makeatletter
\renewcommand\subsection{\@startsection
  {subsection}
  {2}
  {\z@}
  {0.8cm \@plus1ex \@minus .2ex}
  {0.5cm}
  {\normalfont\bfseries}
}
\makeatother

\begin{document}

\title{Two-photon coherent control of femtosecond photoassociation }

\author{Christiane P. Koch}
\email{ckoch@physik.fu-berlin.de}
\affiliation{Institut f\"ur Theoretische Physik,
  Freie Universit\"at Berlin,
  Arnimallee 14, 14195 Berlin, Germany}
\author{Mamadou Ndong}
\affiliation{Institut f\"ur Theoretische Physik,
  Freie Universit\"at Berlin,
  Arnimallee 14, 14195 Berlin, Germany}
\author{Ronnie Kosloff}
\affiliation{Department of Physical Chemistry and
  The Fritz Haber Research Center, 
  The Hebrew University, Jerusalem 91904, Israel}

\maketitle
\renewcommand{\thefootnote}{\fnsymbol{footnote}}

\hrule\medskip

\noindent
Photoassociation with short laser pulses has been proposed as a
technique to create ultracold ground state molecules. 
A broad-band excitation seems the natural choice to drive the series
of excitation and deexcitation steps required to form a molecule in
its vibronic ground state from  two scattering atoms.
First attempts at femtosecond photoassociation were, however, hampered
by the requirement to eliminate the atomic excitation leading to trap
depletion. On the other hand, molecular levels very close to the
atomic transition are to be excited. The broad bandwidth of a
femtosecond laser then appears to be rather an obstacle. 
To overcome the ostensible conflict of driving a narrow
transition by a broad-band laser, 
we suggest a two-photon photoassociation scheme.
In the weak-field regime, a spectral phase pattern can be
employed to eliminate the atomic line.
When the excitation is carried out by more than one photon,
different pathways in the field can be interfered constructively or
destructively. In the strong-field regime, a temporal phase 
can be applied to control dynamic Stark shifts. The atomic transition
is suppressed by choosing a phase which keeps the levels out of
resonance. We derive analytical solutions for atomic two-photon dark
states in both the weak-field and strong-field regime. Two-photon
excitation may thus pave the way toward coherent control of
photoassociation. Ultimately, the success of such a scheme will depend
on the details of the excited electronic states and transition dipole
moments. We explore the possibility of two-photon femtosecond
photoassociation for alkali and alkaline-earth metal dimers and
present a detailed study for the example of calcium. 

\medskip
\hrule
\section{\NoCaseChange{Introduction}}
\label{sec:intro}

Coherent control was conceived to solve the problem of optimizing the
outcome of a chemical reaction \cite{RiceBook,ShapiroBook}. 
Initial work was devoted to photodissociation \cite{BaumertPRL90,KleimanJCP95}. 
In this study, we address the inverse problem of
controlling the free-to-bound transition in photoassociation \cite{JonesRMP06} where
atoms are assembled to a molecule by laser light. 
The underlying principle of coherent control is interference of
different pathways that lead from the initial state to the final
outcome. Typically, in a free-to-bound transition, the relative phase
between the reactants  is not defined \cite{ShapiroBook,ArangoJCP06}: 
If the wavefunction of the initial state can be written as a product
of the wavefunctions of the atoms, the outcome of the binary reaction
cannot be controlled. At very low temperatures, these considerations
do not hold. Ultracold collisions are characterized by threshold
effects \cite{Weiner98,FrancoiseReview}. 
The kinetic energy is very small, and the inner part of the
scattering wavefunction is dominated by the potential energy. As a
result, the colliding pair becomes entangled. Ultracold atoms are
therefore the best candidates for coherent control of a binary
reaction. The formation of molecules in ultracold atomic gases is presently
attracting significant interest \cite{HutsonIRPC06}. 

Previous proposals for short-pulse photoassociation were all based on
one-photon transitions. 
Initially the goal was to optimize the excitation process in
photoassociation by introducing a chirp
\cite{JiriPRA00,ElianePRA04}. In order to produce molecules in their
 ground electronic state, a pump-dump scheme was studied
\cite{MyPRA06a,MyPRA06b,MyJPhysB06}: A first pulse excites two
atoms and creates a molecular wave packet. This is followed by free
dynamics on the electronically excited state bringing the wave packet
to shorter internuclear distances. There a second pulse stabilizes the
product by transferring amplitude to the ground electronic state. The
pump-dump scheme can be supplemented by a third field to engineer the
excited state wave packet dynamics to create favorable conditions for
the dump pulse \cite{MyPRA08}.

\begin{figure}[tb]
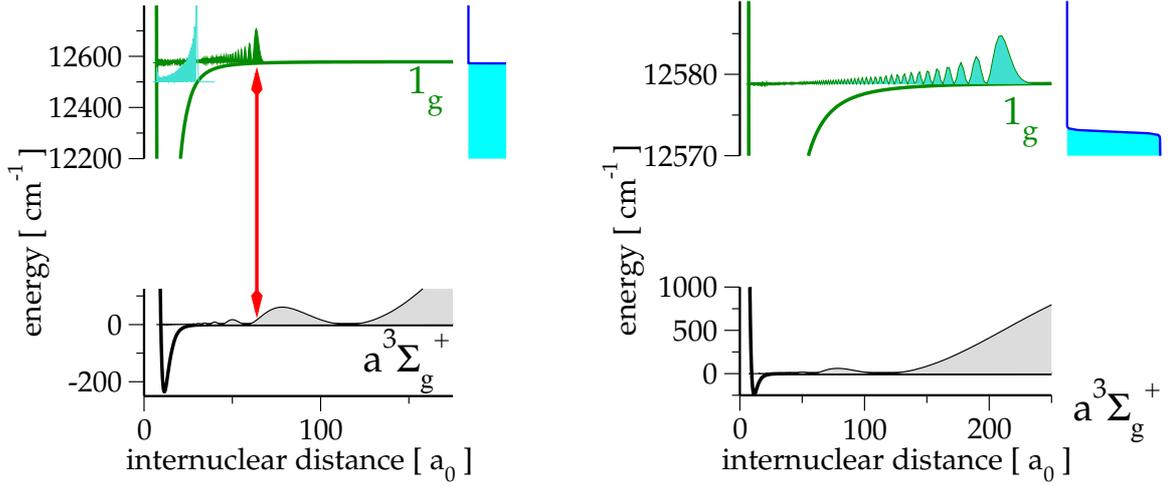

  \centering
  \includegraphics[width=0.45\textwidth]{potD1trip_wvfct}\hspace*{2ex}
  \includegraphics[width=0.45\textwidth]{potD1trip_wvfct2}%
  \caption{%
    In one-photon photoassociation, blocking the
    spectral amplitudes at and close to the
    atomic resonance prevents weakly bound molecular levels with
    large free-bound Franck-Condon factors to be excited (right-hand
    side). The majority of spectral components does not find any
    ground state population to be excited since the probability to
    find two colliding atoms at short internuclear distance is
    tiny (left-hand side, data shown here for Rb$_2$).
    In femtosecond photoassociation employing a one-photon transition, 
    the large majority of spectral components passes through the
    sample without any effect.  
  }
  \label{fig:bandwidth}
\end{figure}

Experiments aiming at femtosecond photoassociation with
a one-photon near-resonant excitation were faced with the obstacle
that weakly bound molecular levels are very close to an atomic resonance.
The challenge is to populate these levels without exciting the atomic
transition. The 
weakly bound molecular levels possess the biggest free-bound
Franck-Condon factors.  However, the atomic transition matrix elements are
several orders of magnitude larger. Excitation of the atomic
transition is followed by spontaneous emission and trap loss, i.e. it
leads to a depletion of the trap
\cite{SalzmannPRA06,BrownPRL06}. Pulses with bandwidths of a few
wavenumbers corresponding to transform-limited pulse durations of a
few picoseconds were proposed as a remedy
\cite{MyPRA06a,MyPRA06b}. However, for such narrow bandwidths, pulse
shaping capabilities have yet to be developed. Experimenters employing
femtosecond pulses with bandwidths of the order of 100$\,$cm$^{-1}$
have resorted to suppressing the spectral amplitude at the atomic
resonance frequency by placing a razor knife into the
Fourier plane of the pulse shaper
\cite{SalzmannPRA06,BrownPRL06,SalzmannPRL08}. 
Due to the finite spectral
resolution, this leads also to the suppression of spectral amplitude
which could excite the molecular levels with the largest free-bound
Franck-Condon factors as illustrated in Fig.~\ref{fig:bandwidth}
(right-hand side). On the other hand, the spectral amplitude from the
peak of the spectrum as well as from just below the cut finds almost no
ground state population to excite since
the probability of finding two colliding atoms at short
internuclear distance is extremely small (left-hand side of
Fig.~\ref{fig:bandwidth}). As a result, only a tiny part of the
spectral amplitude contributes to the photoassociation signal 
\cite{SalzmannPRL08}. 

Here we suggest to employ two-photon transitions for femtosecond
photoassociation. The main idea consists in rendering the atomic
transition dark by applying spectral or temporal phase control while exciting
population into weakly bound molecular levels by two-photon
transitions. The spectral resolution is determined by that 
of the pulse shaper, typically of the order of one wavenumber. This
should allow for the population of weakly bound molecular levels with
the best free-bound Franck-Condon factors.

Two-photon control is based on constructive or
destructive interference of all two-photon pathways adding up to the
two-photon transition frequency
\cite{MeshulachNature98,MeshulachPRA99}.
It allows for the excitation of a very narrow atomic transition by a
broad band femtosecond pulse \cite{MeshulachNature98}. In particular, an
anti-symmetric spectral phase optimizes non-resonant two-photon
population transfer while a symmetric spectral phase yields a dark
pulse \cite{MeshulachNature98}. For weak fields, the two-photon
absorption yield can be calculated within perturbation theory,
allowing for an analytical derivation of suitable phase
functions. At intermediate fields, four-photon pathways additionally
contribute to the two-photon absorption, requiring a higher order
in the perturbation treatment \cite{ChuntonovPRA08,ChuntonovJPB08}. In
both the weak-field and intermediate-field regime, rational pulse shaping
is based on frequency domain arguments 
\cite{MeshulachNature98,MeshulachPRA99,ChuntonovPRA08,ChuntonovJPB08}.
For strong fields, a time-domain picture becomes more adequate since
the dynamics are dominated by time-dependent Stark shifts 
\cite{TralleroPRA05,WeinachtPRL06,TralleroPRA07}.
The atom accumulates a phase due to the dynamic Stark shift.
A suitable control strategy consists in eliminating this phase
by applying a
time-dependent phase function, i.e. by chirping the pulse
\cite{WeinachtPRL06}. This allows for maintaining a $\pi$-pulse
condition despite the levels being strongly Stark shifted. 

These results for coherent control of two-photon absorption in atomic
systems serve as our starting point to derive pulses which are dark at
the atomic transition but optimize two-photon photoassociation,
i.e. population transfer into molecular levels. The shape of the 
pulses yielding a dark resonance of the atoms are derived
analytically: In the weak-field and intermediate-field regimes, the two-photon
spectrum is required to be zero at the atomic transition frequency,
while for strong fields a time-dependent phase allows to maintain an
effective $2\pi$-pulse condition for the atoms. The shaped pulses are tested
numerically for molecule formation. The paper is organized
as follows: An effective
two-state Hamiltonian is derived in Section~\ref{sec:twophoton} by
adiabatically eliminating all off-resonant levels and invoking the
two-photon rotating-wave approximation (RWA). In
Section~\ref{sec:analytical}, conditions for
pulses that are dark at the two-photon atomic transition are derived
analytically. Photoassociation is studied in Sections~\ref{sec:model}
and \ref{sec:numerical} with
Section~\ref{sec:model} reviewing 
possible two-photon 
excitation schemes for alkali and alkaline-earth dimers and
Section~\ref{sec:numerical} presenting the 
numerical study of two-photon photoassociation for calcium.
Section~\ref{sec:concl} concludes.

\section{\NoCaseChange{Theoretical framework for multi-photon
    excitations}} 
\label{sec:twophoton}

We extend the treatment of multi-photon transitions in
atoms
\cite{MeshulachNature98,MeshulachPRA99,ChuntonovPRA08,ChuntonovJPB08,TralleroPRA05,TralleroPRA07}
to include the vibrational degree of freedom, $\Op{R}$, of a diatomic
molecule. To this end, we write the nuclear Hamiltonian for each
electronic state, $\Op{H}_i 
=\Op{T}+V_i(\Op{R})+\omega_i$ (setting $\hbar=1$), i.e. all
potentials, $V_i(\Op{R})$, go to zero asymptotically and the atomic
excitation energies are contained in $\omega_i$. $\Op{T}$ denotes the
vibrational kinetic energy. Assuming the dipole approximation for the
matter-field interaction, the total Hamiltonian is given by
\begin{equation}
  \label{eq:Htot}
  \Op{H}=\Op{H}_0 + \Op{\mu} E(t) \,,
\end{equation}
where $\Op{\mu}$ denotes the dipole operator and $E(t)$ the laser
field with envelope $S(t)$,
$E(t)=\half E_0 S(t) \left( e^{\frac{i}{2}\varphi(t)}e^{-i\omega_L t}
  + c.c.\right)$.
For weak and intermediate fields, the dynamics under the Hamiltonian,
$\Op{H}$, can be solved by perturbation theory.
To lowest order, the two-photon absorption is  given in terms of
the two-photon spectral amplitude on 
resonance \cite{MeshulachNature98,MeshulachPRA99,ChuntonovPRA08},
\begin{equation}
  \label{eq:TPA_pert}
  S_2 \sim \Big| \int_{-\infty}^{+\infty} \tilde E(\omega_{ge}/2+\omega)
  \tilde E(\omega_{ge}/2-\omega) d\omega\Big|^2\,, 
\end{equation}
where $\omega_{ge}$ is  the two-photon resonance frequency  and
$\tilde E(\omega)$ the Fourier transform of
$E(t)$. For an atomic system, conditions for a dark two-photon
resonance are derived analytically by
setting the integral to zero, see below in Section
\ref{subsec:ana_intermed}. 

For strong fields, we follow the derivation of Trallero-Herrero
et. al~\cite{TralleroPRA05,TralleroPRA07}. 
The matter Hamiltonian is expanded in the electronic basis,
$|i\rangle\langle i|$, 
\begin{equation}
  \label{eq:H0}
  \Op{H}_0 = \sum_i\bigg( \Op{T}+V_i(\Op{R})+\omega_i\bigg) |i\rangle\langle i|\,.
\end{equation}
and the electronic states are separated into initial ground state, final
excited state and intermediate states. If the one-photon detunings
of the intermediate states, $\omega_{m,g/e}-\omega_L$, are large compared
to the spectral bandwidth of the laser pulse, $\Delta\omega_L$, the
intermediate states can be adiabatically eliminated
\cite{TralleroPRA05}. Since the electronic excitation energies are
much larger than the vibrational energies, the adiabatic elimination in the
molecular model proceeds equivalently to the atomic case.
Invoking a two-photon RWA, an effective two-state Hamiltonian is
obtained,  
\begin{equation}
  \label{eq:Htwophoton}
  \Op{H}(t) =
  \begin{pmatrix}
    \Op{T} + V_g(\Op{R}) + \omega_g^S(t) & \chi(t)
    e^{-i\varphi(t) }\\
    \chi(t) e^{i\varphi(t)} &  \Op{T} + V_e(\Op{R})+ \Delta_{2P} + \omega_e^S(t)
  \end{pmatrix}\,,
\end{equation}
where $\omega_{g/e}^S(t)$ denotes the dynamic Stark shift of the ground
and excited state, $\chi(t)$ the two-photon coupling and
$\Delta_{2P}$ the two-photon detuning,
$\Delta_{2P}=\omega_{eg} - 2\omega_L$.
The two-photon coupling is given in terms of the one-photon couplings
and one-photon detunings of the ground and excited state to the
intermediate levels, 
\begin{equation}
  \label{eq:chi}
  \chi(t) =-\frac{1}{4}E_0^2 |S(t)|^2 \sum_m
  \frac{\mu_{em}\mu_{mg}}{\omega_{mg}-\omega_L}\,.
\end{equation}
Similarly, the dynamic Stark shifts are obtained as
\begin{equation}
  \label{eq:Stark}
  \omega_i^S(t) = -\half E_0^2 |S(t)|^2 \sum_m |\mu_{mi}|^2
  \frac{\omega_{mi}}{\omega_{mi}^2-\omega_L^2}\,,\quad i=g,e\,.
\end{equation}
The transition dipole matrix elements, $\mu_{ij}$,
and the transitions frequencies, $\omega_{mi}$, 
are in
general $R$-dependent. They tend to a constant value (the atomic
dipole moments and transition frequencies)
at large internuclear distances that are important for
photoassociation. The $R$-dependence is therefore neglected in the
following. 

Eqs.~(\ref{eq:chi}) and (\ref{eq:Stark}) contain couplings between the
intermediate and the ground and excited states, but
couplings between intermediate states are neglected. For very strong
fields, these higher order couplings should be included. 
This type of adiabatic elimination is valid if the Stark shift is
small relative to the energy spacing between the ground and excited states.
The validity of the conditions for adiabatic elimination and two-photon RWA
are easily checked in the atomic case by comparing the dynamics under
the effective Hamiltonian, Eq.~(\ref{eq:Htwophoton}), to those under
the full Hamiltonian, Eq.~(\ref{eq:Htot}), including all intermediate
levels. Possibly the Born-Oppenheimer potential energy surfaces  
in Eq.~(\ref{eq:H0}) show crossings or avoided
crossings in the range of $R$ that is addressed by the pulse.
In such a case, Eq.~(\ref{eq:Htwophoton}) needs to be
extended to include all coupled electronic states.

For strong-field control, it is instructive to choose the rotating
frame such that the phase terms of the coupling in
Eq.~(\ref{eq:Htwophoton}) are expressed as time-dependent frequencies
\footnotemark[3] 
\footnotetext[3]{
  This corresponds to the fact that in the RWA a chirp can be
  expressed either as a phase term of the electric field envelope or
  as instantaneous laser frequency.
  }
and to time-dependently shift the origin of energy \cite{TralleroPRA05}.
The transformed Hamiltonian is written,
\begin{equation}
  \label{eq:Htrans}
  \Op{H}(t) =
  \begin{pmatrix}
    \Op{T} + V_g(\Op{R}) &    \chi(t) \\
    \chi(t) & \Op{T}  + V_e(\Op{R}) + \big(
    \delta_\omega^S(t)+\Delta_{2P}+\dot{\varphi}\big) 
  \end{pmatrix}\,,
\end{equation}
where the differential Stark shift
\begin{equation}
  \label{eq:diffStark}
  \delta_\omega^S(t) = \omega_e^S(t) - \omega_g^S(t)
\end{equation}
is introduced. In the atomic case, strong-field control consists in
locking the phase 
corresponding to the term in parenthesis in Eq.~(\ref{eq:Htrans}),
i.e. in keeping
\begin{equation}
  \label{eq:phaselock}
  \Phi(t) = \int_{-\infty}^t \delta_\omega^S(\tau)d\tau + \Delta_{2P}
  t+ \varphi(t)
\end{equation}
constant. This can be achieved by proper choice of phase of the laser
pulse, $\varphi(t)$, and it
corresponds to maintaining resonance by instantaneously correcting for
the dynamic Stark shift \cite{TralleroPRA07}.

\section{\NoCaseChange{Two-photon dark pulses}}
\label{sec:analytical}

\subsection{Solution in the weak- and intermediate-field regime}
\label{subsec:ana_intermed}

At low field intensities, the population transfer is dominated by the
resonant two-photon pathways. Therefore control strategies are best
understood in the frequency domain \cite{ChuntonovPRA08}. 
Starting from the Hamiltonian in the two-photon RWA,
Eq.~(\ref{eq:Htwophoton}), the condition for a dark pulse at the
atomic transition can be derived analytically. The atomic transition
corresponds to the asymptotic limit of Eq.~(\ref{eq:Htwophoton}),
i.e. to a two-level system. Since the two-photon
coupling, $\chi(t)$, is proportional to the field amplitude squared,
$E_0^2|S(t)|^2$, we look for a spectral phase such that the Fourier
transform of $S(t)^2$,
\[
F(\omega) = \int_{-\infty}^{+\infty} S^2(t) e^{i\omega t} dt =  
\int_{-\infty}^{+\infty} \tilde{S}(\omega') \tilde{S}(\omega-\omega')  d\omega'\, , 
\]
becomes zero at the two-photon resonance, $F(\omega_{eg}) = 0$
($\tilde{S}(\omega)$ is the Fourier transform of the envelope function $S(t)$).
We assume a two-photon resonant pulse, $\omega_L =
\omega_{eg}/2$. Then
\begin{eqnarray}
\label{eq:TFs2-om0}
F(\omega_{eg}) = 0 = F(2\omega_L) &=&
\int_{-\infty}^{+\infty} \tilde{S}(\omega')
\tilde{S}(2\omega_L-\omega')  d\omega' \\
&=&  \int_{-\infty}^{+\infty} \tilde{S}(\omega_L+\omega)
\tilde{S}(\omega_L-\omega)  d\omega \,.  \nonumber
\end{eqnarray}
The condition in Eq.~(\ref{eq:TFs2-om0}) is formally equivalent to
that obtained in second order perturbation theory
\cite{MeshulachNature98,ChuntonovPRA08}.  However, here we have derived
it by considering the effective two-photon Hamiltonian,
Eq.~(\ref{eq:Htwophoton}). 
The condition of Eq.~(\ref{eq:TFs2-om0}) is sufficient for a
two-photon dark pulse as long as the states are not strongly Stark-shifted,
i.e. it holds also at intermediate field intensities. This finding is
rationalized by considering the next order contribution. The
fourth-order terms are classified into on-resonant and near-resonant
terms \cite{ChuntonovPRA08}. The dominant fourth-order term is the one
that includes the resonant second-order term and is therefore
canceled by the same condition. 

We assume the pulse shape to be Gaussian. Then $\tilde S(\omega)=A_0
e^{-\frac{(\omega-\omega_L)^2}{2\sigma_\omega^2}}e^{i\phi(\omega)}$
with $\sigma_\omega$ the spectral bandwidth, 
and  the two-photon Fourier transform becomes 
\[
F(2\omega_L) = A_0^2 \int_{-\infty}^{+\infty}
e^{-\frac{\omega^2}{\sigma_\omega^2}}e^{i\big(\phi(\omega_L+\omega)+\phi(\omega_L-\omega)
  \big)}
d\omega \,.
\]
We now need to choose the spectral phase $\phi(\omega)$ such as to
render $F(2\omega_L)$ zero. This can be achieved 
by a step of the phase $\phi(\omega)$
 by $\pi$ \cite{MeshulachNature98,MeshulachPRA99}.

We derive here the position of this step, $\omega_s$. Let us assume
that $\omega_s > \omega_L$, and denote
$\Delta\omega_s=\omega_s-\omega_L$. Then
\begin{equation}
\phi(\omega_L+\omega)+\phi(\omega_L-\omega) = \left \{
\begin{array}{l}
0 \qquad  \mathrm{if} \qquad \omega \geq \Delta\omega_s \,, \\
0  \qquad  \mathrm{if} \qquad \omega \leq -\Delta\omega_s \,, \\
\pi \qquad \mathrm{if} \qquad -\Delta\omega_s < \omega <
\Delta\omega_s \,.
\end{array}
\right. 
\label{eq:phase}
\end{equation}
The $\pi$-step implies a change of sign in the integral, and 
the dark pulse condition is rewritten accordingly,
\begin{eqnarray*}
F(2\omega_L) / A_0^2 = 0 &=& 
\int_{-\infty}^{-\Delta\omega_s} e^{-\frac{\omega^2}{\sigma_\omega^2}}
d\omega + 
\int_{\Delta\omega_s}^{+\infty} e^{-\frac{\omega^2}{\sigma_\omega^2}} d\omega -
\int_{-\Delta\omega_s}^{\Delta\omega_s}
e^{-\frac{\omega^2}{\sigma_\omega^2}} d\omega \,,\\
&=& 2 \int_{\Delta\omega_s}^{+\infty}
e^{-\frac{\omega^2}{\sigma_\omega^2}} d\omega -
2 \int_0^{\Delta\omega_s} e^{-\frac{\omega^2}{\sigma_\omega^2}} d\omega \,.
\end{eqnarray*}
We recognize the error function up to a factor and obtain
\begin{equation}
  \label{eq:result}
  \Fkt{Erf} \left(\frac{\Delta\omega_s}{\sigma_\omega} \right) = \half \,,
\end{equation}
which determines $\Delta\omega_s$ and hence the step position, $\omega_s$.
That is, if one chooses $\Delta\omega_s$ such that
$\Fkt{Erf}(\sigma\Delta\omega_s) = 1/2$, the two-photon spectrum is zero at
the two-photon transition frequency,
$F(2\omega_L)=0$, and a dark pulse is obtained.
Interestingly we can also combine the conditions of a symmetrical
function and a phase step. This is achieved by defining two phase
steps symmetrical around $\omega_L$, $\omega_L \pm \omega_s$.
Following a derivation along the
same lines as outlined above, the identical condition for $\omega_s$  is
obtained. 

\subsection{Solution in the strong-field regime}
\label{subsec:ana_strong}

At high intensities, the effective Hamiltonian varies during the
pulse. A time-domain picture is therefore best adapted to develop
control strategies. 
Population inversion in the effective Hamiltonian can be achieved by
adjusting the temporal phase of the laser such as to maintain a
$\pi$-pulse condition \cite{TralleroPRA05,WeinachtPRL06}. Here we turn
this strategy upside-down to achieve a dark pulse, i.e. we ask for the
$2\pi$-pulse condition. We restrict the notation to the atomic case
for simplicity. The Hamiltonian, Eq.~(\ref{eq:Htrans}),
can be transformed such that the diagonal elements
are zero \cite{TralleroPRA07},
\begin{equation}
  \label{eq:Htimedep}
\Op{H}(t) =
\begin{pmatrix}
  0 & \chi(t) e^{-i\Phi(t)} \\
  \chi(t) e^{i\Phi(t)} & 0 
\end{pmatrix} \,.  
\end{equation}
Evaluating Rabi's formula for the final state amplitude
leads to the condition
\begin{equation}
  \label{eq:2pi}
  \left|\int_0^{t_f} \chi(t) e^{i\Phi(t)} dt \right| = 2k\pi \,,\quad
k=0,1,\ldots
\end{equation}
for zero population transfer to the excited state.
We again assume a two-photon resonant pulse,
$\omega_{ge}=2\omega_L$. The condition, Eq.~(\ref{eq:2pi}), is
fulfilled given the phase $\Phi(t)$ obeys
\begin{equation}
  \label{eq:phasestrong}
\Phi(t) = \left\{
    \begin{array}{l}
      \pi \qquad \mathrm{if} \qquad t_0-\Delta t_x < t < t_0 + \Delta
      t_x \,,\\
      0 \qquad \mathrm{elsewhere}\,.
    \end{array}
    \right.
\end{equation}
This corresponds to the following temporal laser phase
\begin{equation}
  \label{eq:phasestrongfield}
  \varphi(t) =  \left\{
    \begin{array}{ll}
      -\pi-\int_0^t \delta_\omega^S(\tau) d\tau
      \qquad & \mathrm{if} \qquad t_0-\Delta t_x < t < t_0 + \Delta
      t_x \\
      -\int_0^t \delta_\omega^S(\tau) d\tau \qquad &\mathrm{elsewhere}
    \end{array}
    \right.
\end{equation}
For a Gaussian pulse, the time-dependence of the differential Stark
shift is also Gaussian, and $\varphi(t)$ can be evaluated
analytically. Inserting the definition of $\Phi(t)$,
Eq.~(\ref{eq:phaselock}), into 
the condition, Eq.~(\ref{eq:2pi}), and evaluating 
the integrals yields a condition for $\Delta t_x$,
\begin{equation}
  \label{eq:resultstrongfield}
  \Fkt{Erf}\left(\frac{\Delta t_x}{\sigma} \right) = \half \,.
\end{equation}
A pulse with the temporal phase defined by
Eqs.~(\ref{eq:phasestrongfield}) and (\ref{eq:resultstrongfield})
suppresses population transfer by keeping the two levels out of
resonance. 

Such a solution can also be obtained by applying local
control~\cite{k89} to the effective Hamiltonian, Eq.~(\ref{eq:Htimedep}). 
Local control theory allows to carry the analysis over to the full
molecular problem. We can seek conditions which eliminate
atomic two-photon transitions while maximizing the
molecular population transfer. To this end, we define an operator to
be minimized,
\begin{equation}
  \label{eq:projat}
  \Op{P}_{at} = \int_{R_0}^{+\infty} dR |e\rangle\langle e|
  \,,
\end{equation}
which projects onto the atomic levels, $|e\rangle$ denotes the excited
electronic state. $R_0$ is a distance which is larger than the bond length
of the last bound level of the excited state potential. 
The objective to be maximized is defined by the projection onto the
molecular levels for photoassociation,
\begin{equation}
  \label{eq:projmol}
  \Op{P}_{mol} = \sum_i |\varphi_i^e\rangle\langle \varphi_i^e|\,,
\end{equation}
where $|\varphi_i^e\rangle$ denote vibrational eigenstates of the
electronically excited state.
In local control theory, we seek the phase of the field such that
\begin{equation}
  \label{eq:localcontrol}
  \frac{d}{dt}\langle\Op{P}_{at}\rangle =0 \quad\quad
  \frac{d}{dt}\langle \Op{P}_{mol}\rangle > 0\,,
\end{equation}
i.e. atomic transitions are suppressed while the molecular population 
is monotonically increasing \cite{AllonJCP93}. Eq.~(\ref{eq:phasestrongfield})
is a specific analytical example of local control. The first part of the pulse
up to time $t \le t_0 -\Delta t_x$ builds up excited state population
and a phase relation between the ground and excited levels.  
At this instant in time the phase of the field jumps by $\pi$ so that
it becomes perpendicular to the phase of the instantaneous transition
dipole, thus eliminating additional population transfer.  The last
part of the pulse restores the original population.

\section{\NoCaseChange{Two-photon excitation schemes for
    photoassociation}} 
\label{sec:model}

Our goal is to populate molecular levels in a long-range potential
while suppressing excitation of the atomic transition. We will
restrict the discussion to  homonuclear dimers where potentials with
$1/R^5$ behavior occur below the $S+D$ asymptote\footnotemark[1]
\footnotetext[1]{One may
  also consider transitions into potentials correlating to an $S+P$
  asymptote. While this is two-photon forbidden for atoms, it might be
  two-photon allowed at short internuclear distances. However, in that
  case photoassociation will occur only at distances below 20 Bohr.
  This corresponds to large detunings, and a two-photon control
  scheme is then not required.}.
In heteronuclear dimers, the potentials go as $1/R^6$, and the
improvement of two-photon over one-photon femtosecond photoassociation
is expected 
to be less significant. Alkali and alkaline earth species 
where cooling and trapping of atoms is well established are
considered. Carrier frequencies in the near-infrared are assumed
throughout.

\subsection{Two-photon transitions in alkali dimers}

A one-photon transition cannot be controlled by interference between
different photon pathways. A prerequisite for our scheme is therefore 
that no one-photon resonances be contained within the pulse
spectral bandwidth. This excludes both potassium and rubidium as
possible candidates. Of the remaining  alkalis, only caesium 
shows a two-photon resonance between the ground state and a $d$-level
in the near-infrared, cf. Table~\ref{tab:trans}.

\subsection{Two-photon transitions in alkaline earth dimers}

\begin{figure}[tb]
  \centering
  \includegraphics[width=0.7\textwidth]{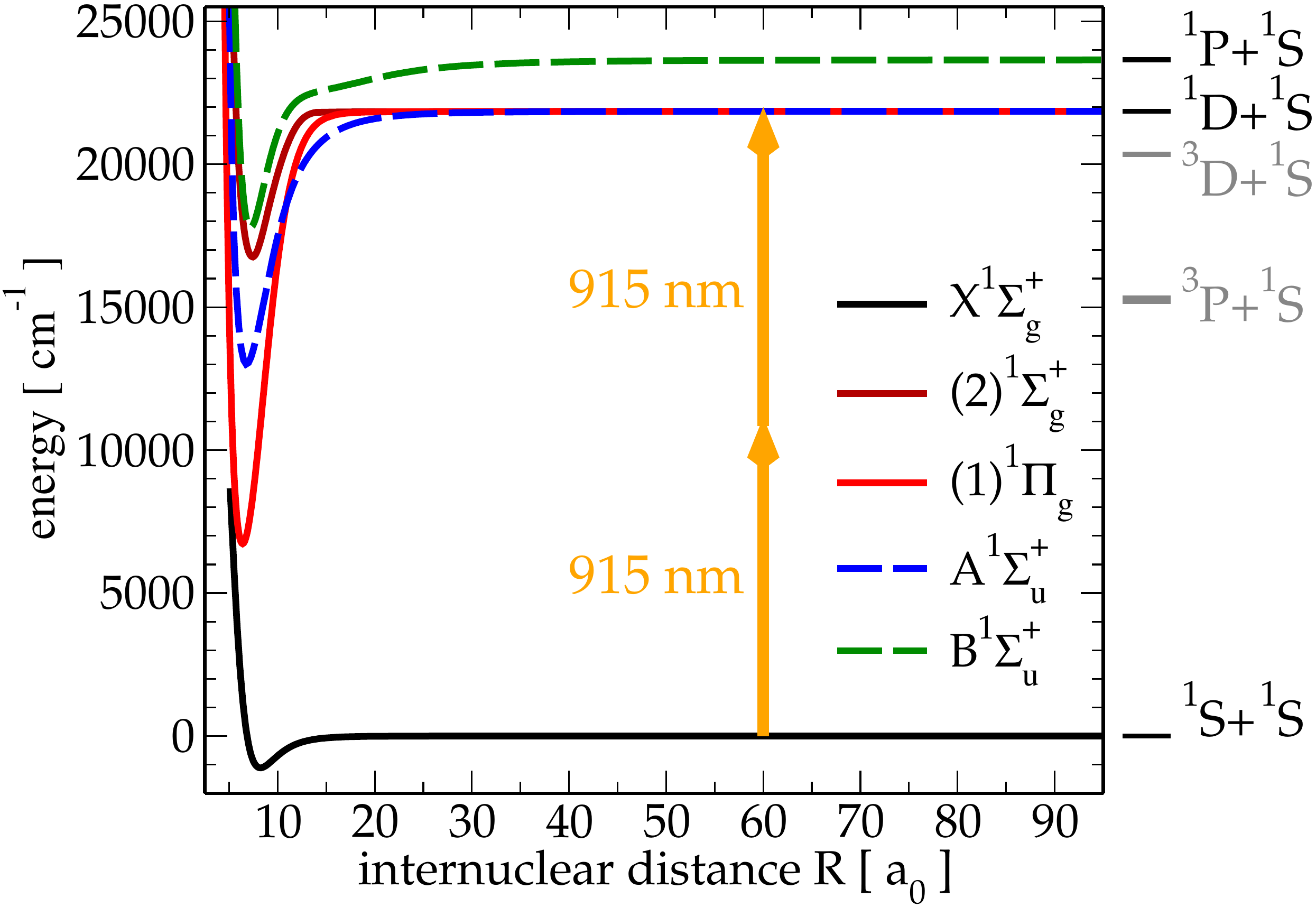}
  \caption{Two-photon photoassociation of two atoms colliding over the
    $X^1\Sigma_g^+$ ground state into the $(1)^1\Pi_g$-state or the
    $(2)^1\Sigma_g^+$-state correlating to the $s+d$ asymptote (solid
    lines).
    The dipole coupling is provided by the states with one-photon
    allowed transitions from the ground state such as the
    $A^1\Sigma_u^+$-state or the $B^1\Sigma_u^+$-state (dashed lines).
    The example shown here is Ca$_2$.}
  \label{fig:twophotonCa2}
\end{figure}

Of the alkaline earth atoms, calcium and strontium show two-photon
resonances at the desired wavelength, cf. Table~\ref{tab:trans}, and
in magnesium a near-IR four-photon transition to a $d$ level is
possible. 
In the following we will study calcium since potential curves for a
large number of electronically excited states are known with
sufficient accuracy \cite{RobertMolPhys06}.\footnotemark[2]
\footnotetext[2]{We expect that our
reasoning can be carried over to strontium with only slight
modifications due to the similar electronic structure. }
The two-photon excitation
scheme is shown in Fig.~\ref{fig:twophotonCa2}: Two atoms interacting via
the $X^1\Sigma_g^+$-state are excited into bound levels of the 
$(1)^1\Pi_g$-state or the $(2)^1\Sigma_g^+$-state that both correlate
to the $^1S+^1D$ asymptote. Note that the $^1D$ term of calcium occurs
energetically below the  $^1P^0$ term. 
The dipole coupling is provided by the states
with one-photon allowed transitions from the ground state,
i.e. the $A^1\Sigma_u^+$, $A'^1\Pi_u$, $B^1\Sigma_u^+$ states and
further electronic states at higher energies. The effective two-photon
coupling,  cf. Eq.~(\ref{eq:chi}),
is determined by the  dipole matrix elements of the one-photon allowed
transitions  and by the one-photon detunings. 
Compared to the atomic two-photon transitions 
studied in caesium \cite{MeshulachNature98} and sodium
\cite{WeinachtPRL06,TralleroPRA07,ChuntonovPRA08,ChuntonovJPB08}, the
one-photon detunings in the alkaline earths are much larger. 
Higher laser intensities will therefore be required.

\begin{figure}[ht]
  \centering
  \includegraphics[width=0.7\textwidth]{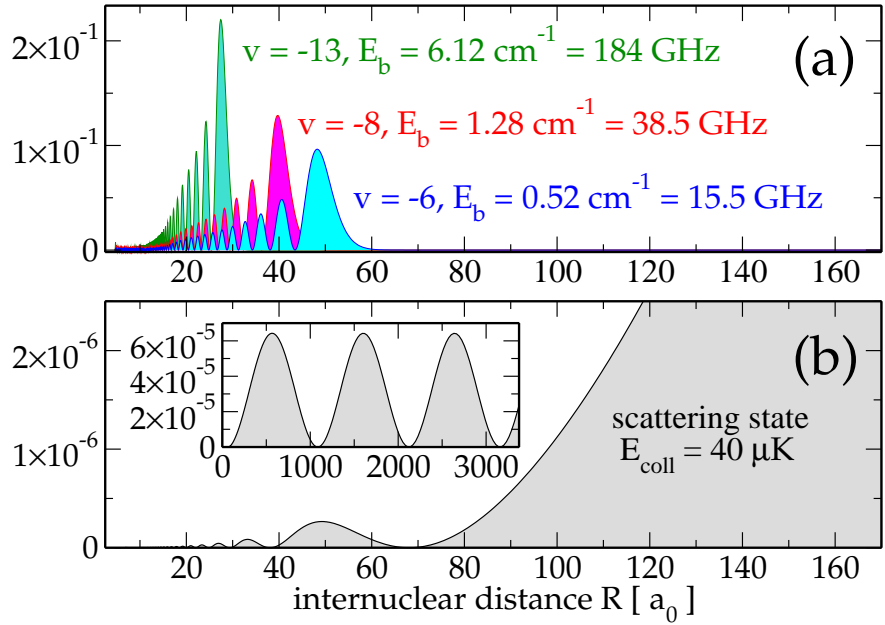}
  \caption{
    Weakly bound
    vibrational wavefunctions of the $(1)^1\Pi_g$ excited state for
    three different binding energies relevant for photoassociation 
    (a) and scattering wavefunction of two ground state calcium atoms (b). 
  }
  \label{fig:vib}
\end{figure}

Fig.~\ref{fig:vib} demonstrates
the existence of long-range vibrational wavefunctions in a potential
with $1/R^5$ behavior for large internuclear distances $R$.
Vibrational wavefunctions of the $(1)^1\Pi_g$
excited state potential are plotted in Fig.~\ref{fig:vib}a for three
different binding energies. The wavefunctions need to be compared to 
the initial scattering state, shown in Fig.~\ref{fig:vib}b for
two ground state atoms colliding with approximately 40$\,\mu$K: Large
free-bound Franck-Condon factors are due to sufficient probability
density at large $R$ in 
the excited state vibrational wavefunctions. In the following
population shall be transferred from the initial ground state atomic
density into these vibrational wavefunctions close to the
excited-state dissociation limit. The efficiency will be determined by
the width of the dark atomic resonance, i.e. by how close to the
dissociation limit excitation into molecular levels can occur.

\section{\NoCaseChange{Two-photon photoassociation in calcium}} 
\label{sec:numerical}

Photoassociation of two calcium atoms is studied by numerically
solving the time-dependent Schr\"odinger equation for the effective
two-photon Hamiltonian, Eq.~(\ref{eq:Htwophoton}), with a Chebychev
propagator \cite{RonnieReview94}. The Hamiltonian is represented on a
grid using an adaptive grid step size
\cite{SlavaJCP99,WillnerJCP04,ShimshonCPL06}. The two-photon couplings
and dynamic Stark shifts are calculated from the atomic transition
dipole matrix elements and frequencies found in Ref. \cite{NIST}. The
potentials are gathered from Ref. \cite{RobertMolPhys06}. An initial
state with a scattering energy corresponding to 40$\,\mu$K,
cf. Fig.~\ref{fig:vib}b, is chosen.

\begin{figure}[ht]
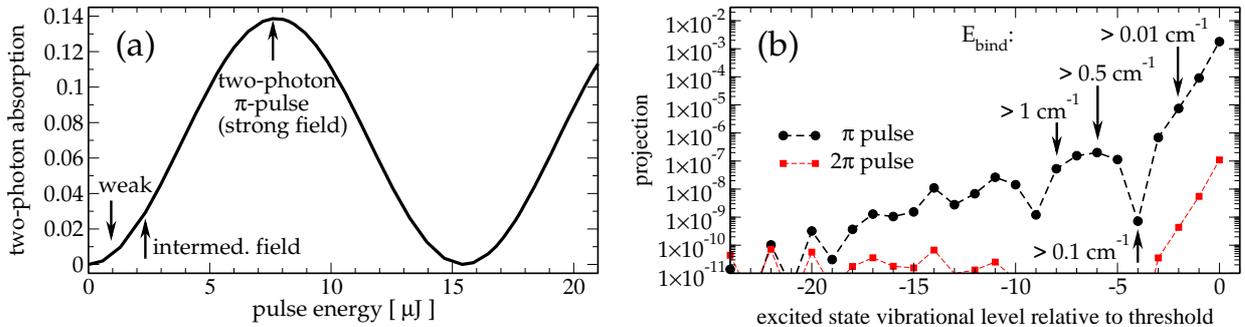

  \centering
  \includegraphics[width=0.49\textwidth]{2Ppipulse}\hspace*{2ex}
  \includegraphics[width=0.49\textwidth]{vibdist}
  \caption{(a) The excited state population shows 
    Rabi oscillations  as the pulse energy is increased. The arrows
    indicate the pulsed energies which are employed for shaped pulses.
    (b) The projection of the final excited state wave packet onto the
    molecular levels, i.e. the excited state vibrational
    distribution after the pulse for a two-photon $\pi$-pulse and a
    two-photon $2\pi$-pulse. The arrows indicate the range of binding
    energies relevant in the discussion of the results for shaped
    pulses. }
  \label{fig:TL}
\end{figure}

Fig.~\ref{fig:TL} presents calculations with transform-limited pulses
of 100$\,$ fs full-width 
half-maximum and a central wavelength of 915$\,$nm.
They serve as reference point for the shaped pulse
calculations discussed below. 
Fig.~\ref{fig:TL}a shows the final excited state population including
both atoms and molecules as a function of pulse energy.
The arrows
in Fig.~\ref{fig:TL}a indicate pulse energies in the weak,
intermediate and strong field regime  that are employed in
shaped pulse calculations. Rabi
oscillations are observed as the pulse energy is increased. The
contrast of the oscillations is reduced from 100\% to 14\%. This is
a clear sign of the dynamic Stark shift for strong fields.
For strong (intermediate) field, the maximum Stark shift,
Eq.~(\ref{eq:Stark}), 
amounts to $104\,$cm$^{-1}$ ($32\,$cm$^{-1}$)
for the ground state and to $247\,$cm$^{-1}$ ($76\,$cm$^{-1}$)
for the excited state.
The weak field corresponds to a pulse
energy of $0.94\,\mu$J and yields an excited state population of
$5.0\times 10^{-3}$, the intermediate field to $2.4\,\mu$J and
$P_{exc}(t_{final})=2.9\times 10^{-2}$, and the strong field to
$7.6\,\mu$J and $P_{exc}(t_{final})=1.4\times 10^{-1}$. 

The projection of the final wave packet onto the
molecular levels of the excited state, i.e. the excited state vibrational
distribution after the pulse, is shown in Fig.~\ref{fig:TL}b for 
transform-limited two-photon $\pi$
and $2\pi$-pulses. For a $\pi$-pulse, the population of the last bound
level amounts to $1.8\times 10^{-3}$ compared to an atomic population
of $1.4\times 10^{-1}$.  For comparison, one-photon photoassociation
with a 10$\,$ps-pulse detuned by $4\,$cm$^{-1}$ achieves a population
transfer into molecular levels on the order of  $10^{-4}$
which corresponds, depending on the trap conditions, to
1-10 molecules per pulse \cite{MyPRA06b,MyJPhysB06}.
The arrows in  Fig.~\ref{fig:TL}b indicate
the binding energies of a few weakly bound levels. 
Due to their smaller free-bound Franck-Condon
factors, deeper molecular levels are populated significantly less
than the last bound level. The minimum of the
excited state vibrational distribution at $v=-4$ is due to the node of
the scattering wavefunction at $R\sim 68\,$a$_0$,
cf. Fig.~\ref{fig:vib}b. 

\begin{figure}[ht]
  \centering
  \includegraphics[width=0.9\textwidth]{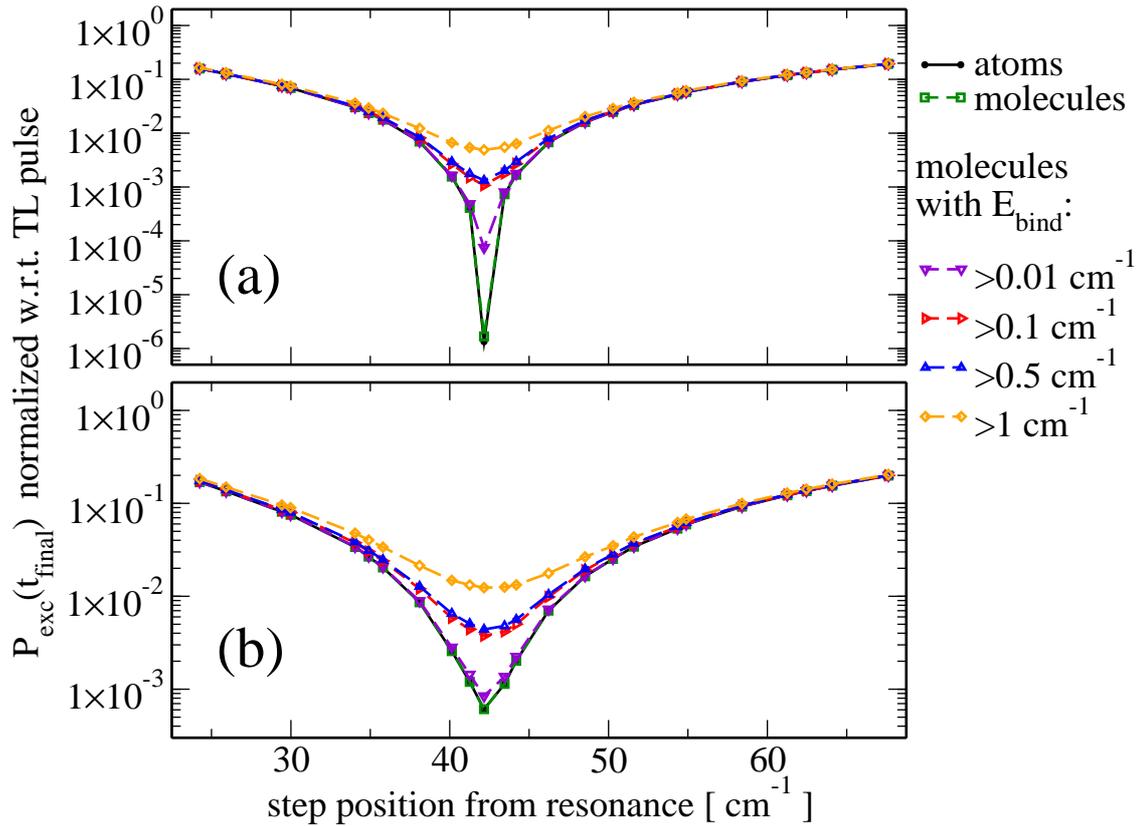}  
  \caption{Frequency-domain control with pulses shaped
    according to Eq.~(\ref{eq:phase}):
    The excited state population of atoms (filled circles) and of
    molecules (open symbols), normalized with respect to the results
    obtained for a transform-limited pulse, is shown as a function of
    the $\pi$-step position for (a) weak and (b) intermediate fields.
    Population transfer can be strongly suppressed, by $10^{-6}$, for
    the weak field (a). For the intermediate field (b), the dynamic Stark
    shift sets in yielding a maximum suppression on the order of
    $10^{-3}$.     
    When all molecular levels are considered, the last two levels (that
    have binding energies of less than 0.01$\,$cm$^{-1}$) dominate and no
    difference between atoms and molecules is observed. For 
    increasing binding energy of the molecular levels, and hence
    detuning from the atomic  resonance,  the two-photon transition
    becomes less dark.
  }
  \label{fig:w-control}
\end{figure}

Frequency-domain control is studied for weak and intermediate fields
in Fig.~\ref{fig:w-control}. The shaped pulses are obtained by
applying a spectral phase function consisting of a $\pi$-step to a
transform-limited 100$\,$fs-pulse, cf. Eq.~(\ref{eq:phase}). The
position of the $\pi$-step is varied. 
Atomic population transfer  (black filled circles)
can be strongly suppressed, by $10^{-6}$, for
the weak field, see Fig.~\ref{fig:w-control}a.
For the intermediate field,  Fig.~\ref{fig:w-control}b, the dynamic
Stark shift moves the levels during the pulse. The dark-pulse
condition taking into account only the static resonance frequency,
Eq.~(\ref{eq:TFs2-om0}), is then not sufficient anymore to strongly 
suppress atomic population transfer. The maximum
suppression,  i.e. the minimum of the final excited state
population, is therefore reduced to be on the order of $10^{-3}$.     
It is observed at the position of the $\pi$-step corresponding to
Eq.~(\ref{eq:result}), detuned by $42.2\,$cm$^{-1}$ from the carrier
frequency for both weak and intermediate field.

When population transfer into all molecular levels is considered,
no difference between atoms and molecules is observed, compare the
black filled circles and open green squares in
Fig.~\ref{fig:w-control}. In this case the last 
two levels which have the largest free-bound
Franck-Condon factors  dominate the molecular population transfer.
Since their binding energy is very small, 
less than 0.01$\,$cm$^{-1}$, there is almost no detuning with respect
to the atomic two-photon resonance, and the dark-state condition
applies to the atomic transition and transitions into the last two
bound levels alike. As the binding energy of the molecular levels and
hence the detuning from the atomic resonance increases, the 
two-photon dark state condition which is defined at the atomic
resonance becomes less applicable. Subsequently,
molecular two-photon transitions become less dark, cf. open
triangles and diamonds in Fig.~\ref{fig:w-control}. 
Atomic transitions are suppressed by about four orders of magnitude
more then transitions into molecular levels with binding energies
larger than $1\,$cm$^{-1}$ for the weak field. This needs to be compared to the
difference in atomic and molecular transition matrix elements. For
weakly bound levels, it amounts to four to five orders of
magnitude. The suppression of the atomic transition in
Fig.~\ref{fig:w-control}a implies therefore that about the same number of atoms
and of molecules bound by more than  $1\,$cm$^{-1}$ will be excited. The
excited atoms will be lost. However, this loss is sufficiently
small, and the trap is not depleted. Due to the high repetition rate
in femtosecond experiments, many photoassociation pulses can be
applied before any significant depletion of the trap occurs. At
intermediate field strength, the atomic transition relative to
transitions into molecular levels with binding energies larger than
$1\,$cm$^{-1}$ is suppressed by only two orders of magnitude. In that
case, loss of atoms from the trap becomes significant, and 
the control strategy of applying a spectral $\pi$-step phase function
is not sufficient to accumulate molecules by applying many
photoassociation pulses.

\begin{figure}[ht]
  \centering
  \includegraphics[width=0.9\textwidth]{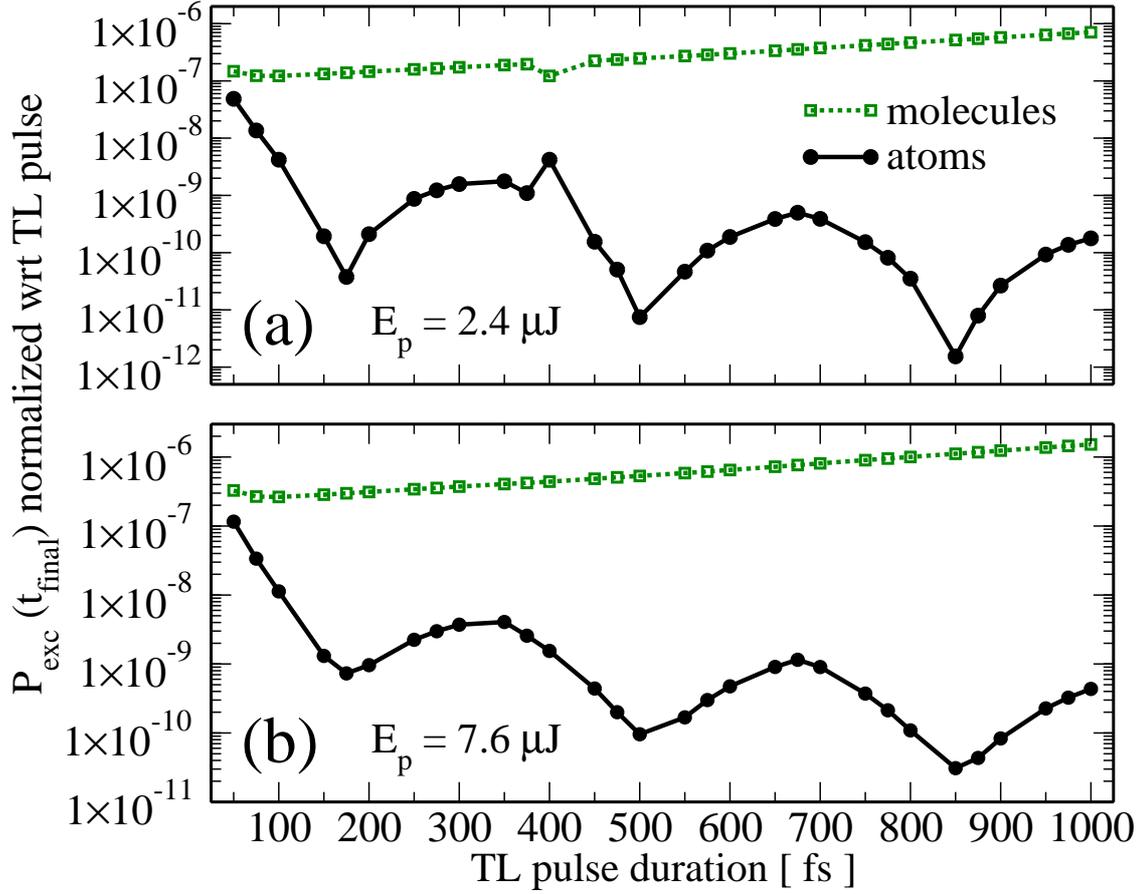}  
  \caption{Time-domain control with pulses shaped according to
    Eq.~(\ref{eq:phasestrongfield}): 
    The excited state population of atoms (filled circles) and of
    molecules (open symbols), normalized with respect to the results
    obtained for a transform-limited pulse, is shown as a function of
    the duration
    of the corresponding transform-limited pulse for intermediate (a)
    and strong field (b). Two-photon transitions for atoms are
    more strongly suppressed than for  molecules. However, the
    suppression of the molecular transitions is too strong to achieve
    any significant photoassociation yield. For weakly bound levels no
    difference in the suppression with increasing binding energy is
    observed, i.e. the levels have very similar dynamic Stark shifts.}  
  \label{fig:t-control}
\end{figure}

For intermediate and strong fields, pulse shaping in the time-domain
is more appropriate since it allows to correct for the dynamic
Stark shift. Time-domain control is studied in
Fig.~\ref{fig:t-control}. The shaped pulses are obtained by
chirping the pulse according to Eq.~(\ref{eq:phasestrongfield}) with
the chirp time interval, $\Delta t_x$, 
determined from Eq.~(\ref{eq:resultstrongfield}). The final excited
state population, normalized with respect to the transform limited
case, is compared for atoms and molecules in Fig.~\ref{fig:t-control} for
intermediate (a) and strong (b) fields. It is plotted for increasing
duration of the transform-limited pulse from which the chirped pulse
is generated, keeping the pulse energy constant. For time-domain
control, two-photon transitions for 
atoms are clearly more strongly suppressed than for molecules. The
ratio of suppression of atoms versus molecules varies as a function of the
transform-limited pulse duration between one and five orders of
magnitude demonstrating a large amount of control.
The overall increase of the difference between atoms and
molecules with increasing pulse duration is rationalized in terms of
the control strategy: The dynamic Stark shift together with the
two-photon detuning leads to the accumulation of a phase in the atoms
and molecules. It is countered by the laser phase $\varphi(t)$,
cf. Eq.~(\ref{eq:phaselock}). The difference between atoms and
molecules becomes more perceptible for longer phase accumulation
times. 
The modulations seem to be caused by a periodic accumulated phase of the 
instantaneous transition dipole. More analysis using local control theory is required.
Overall, however, the suppression of the molecular transitions is much
too strong to achieve any significant photoassociation
yield. Analysing the molecular transitions for the weakly bound levels
relevant in photoassociation, no dependence on binding
energy is observed. Transitions into levels bound by $1\,$cm$^{-1}$ or
more are equally suppressed as transitions into the last two bound
levels. This is not surprising since the levels have a very similar
dynamic Stark shift. Differences arise mostly from the different
two-photon detunings, and they are much smaller than the
dynamic Stark shift.
It is hence not enough to enforce a minimization of the
atomic transition probability.  The
molecular transition probabilities need to be maximized explicitly,
e.g. employing local control and  Eq.~(\ref{eq:localcontrol}).

\section{\NoCaseChange{Summary and conclusions}}
\label{sec:concl}

We have studied coherent control of femtosecond photoassociation
employing two-photon transitions. The goal of our study was to
identify pulses which populate molecular levels close to the atomic
transition without exciting the atoms themselves. An effective
two-photon Hamiltonian was derived by extending previous work on atoms
\cite{TralleroPRA05,WeinachtPRL06}
by the vibrational degree of freedom.
In the weak-field to
intermediate-field regime, frequency-domain control was pursued 
based on applying a spectral phase function. This leads to
constructive or destructive interference between all pathways
contributing to the two-photon absorption
\cite{MeshulachNature98,MeshulachPRA99,ChuntonovPRA08,ChuntonovJPB08}.   
Time-domain control is more
appropriate for the intermediate-field to strong-field regime where
the dynamics are dominated by dynamic Stark shifts
\cite{TralleroPRA05,WeinachtPRL06}. A temporal phase 
function allows to lock the atomic phase to that of the laser, keeping
the atoms on or out of resonance. 
For both regimes, we derived analytical conditions  to 
construct pulses which leave the atomic
two-photon transition dark. These pulses were tested in a numerical
study of photoassociation in calcium. An excited-state potential
correlating to the $^1S+^1D$ asymptote  was chosen. It 
shows a $1/R^5$ behavior at long range and supports weakly bound
molecular levels with large free-bound Franck-Condon factors required
for efficient photoassociation.  

In the weak-field regime, applying a $\pi$-phase step reduces the
atomic transition probability by a factor of $10^{-6}$. Molecular
levels with binding energies of about $1\,$cm$^{-1}$ are sufficiently
detuned from the atomic two-photon resonance that the dark condition
does not apply. Transitions into these levels are suppressed by only a
factor of $10^{-2}$. The resulting four orders of magnitude difference
between atomic and molecular transitions is sufficient to counter the
different excitation probabilities for atoms and molecular levels. It
hence  allows for employing
femtosecond pulses without depleting the trap due to excitation of
atoms. However, weak-field control implies that absolute
excitation probabilities are very small. 

In the intermediate-field regime, the Stark shift becomes large enough
to move the levels during the pulse. The spectral dark pulse condition taking
into account only the static two-photon resonance is then not sufficient to
suppress atomic transitions. A reduction of the atomic transition
probability by merely a factor of $10^{-3}$ was found. Applying a
femtosecond pulse will then excite significantly more atoms than
molecules. Since the excited atoms are lost, trap depletion is
expected to set in after a few pulse cycles. 

Time-domain control based on maintaining a $2\pi$-pulse condition for
the atomic transition is applicable for both intermediate and strong
fields. The ratio of suppression of atomic versus molecular
transitions depends on the duration of the transform-limited pulses
from which the shaped pulses are generated. It reaches up to five orders
of magnitude. This finding is rationalized in
terms of accumulation of different phases for atoms and molecules. 
In absolute numbers, however, the molecular transitions themselves are
strongly suppressed. A significant photoassociation yield can
therefore not be achieved by the simple control strategy based on the
analytical result for the atomic two-photon excitation. Local control
should be employed to maximize the molecular transition probabilities
while keeping the atomic line dark. This is beyond the scope of the
present study. 

In conclusion, an impressive amount of control yielding several orders
of magnitude difference between atomic and molecular transitions 
could be demonstrated with both spectral and temporal phase
shaping. However, the two-photon photoassociation 
efficiencies need to be improved in terms of absolute numbers of
molecules per pulse to yield a feasible photoassociation scheme. In
the strong-field regime, local control offers a means for explicit
maximization of molecular transition probabilities. 
Moreover, our discussion could be extended from two-photon to
three-photon transitions. Then, excited-state potentials correlating to
$S+P$ asymptotes and showing a $1/R^3$ behavior at long range could be
employed. A $1/R^3$ long-range potential supports spatially more
extended vibrational wavefunctions than a $1/R^5$ potential and
exhibits an overall larger density of vibrational levels. Therefore,
the efficiency of a shaped pulse which allows to excite weakly bound
molecular levels in a three-photon transition while keeping the atomic
resonance dark is expected to by far exceed that of the two-photon scheme.

\section*{Acknowledgements}
We wish to thank Zohar Amitay and Yaron Silberberg for fruitful discussions.
Financial support from the Deutsche Forschungsgemeinschaft through the
Emmy Noether programme and SFB 450 is
gratefully acknowledged. 

\clearpage

\bibliographystyle{fd}
\bibliography{twophoton}

\clearpage
\begin{table}[tb]
  \centering
  \begin{tabular}{| c | c | c |}
    \hline
    atom & transition & wavelength \\ \hline
    Cs & $\quad 6s \to 7d \quad$ & $\quad 2\times 768\,$nm$\quad$ \\\hline
    Mg & $\quad 3s^2 \to 3s3d\quad$ & $\quad 4\times 862\,$nm$\quad$ \\
    Ca & $\quad 4s^2 \to 3d4s\quad$ & $\quad 2\times 915\,$nm$\quad$ \\
    Sr & $\quad 5s^2 \to 4d5s\quad$ & $\quad 2\times 992\,$nm$\quad$ \\ \hline
  \end{tabular}
  \caption{Two-photon transitions which are candidates for two-photon
    photoassociation in the near-infrared for alkali and alkaline earth atoms}
  \label{tab:trans}
\end{table}

\clearpage

\begin{list}{}{\leftmargin 2cm \labelwidth 1.5cm \labelsep 0.5cm}
\item[\bf Fig. 1] 
    In one-photon photoassociation, blocking the
    spectral amplitudes at and close to the
    atomic resonance prevents weakly bound molecular levels with
    large free-bound Franck-Condon factors to be excited (right-hand
    side). The majority of spectral components does not find any
    ground state population to be excited since the probability to
    find two colliding atoms at short internuclear distance is
    tiny (left-hand side, data shown here for Rb$_2$).
    In femtosecond photoassociation employing a one-photon transition, 
    the large majority of spectral components passes through the
    sample without any effect.  
\item[\bf Fig. 2] Two-photon photoassociation of two atoms colliding over the
    $X^1\Sigma_g^+$ ground state into the $(1)^1\Pi_g$-state or the
    $(2)^1\Sigma_g^+$-state correlating to the $s+d$ asymptote (solid
    lines).
    The dipole coupling is provided by the states with one-photon
    allowed transitions from the ground state such as the
    $A^1\Sigma_u^+$-state or the $B^1\Sigma_u^+$-state (dashed lines).
    The example shown here is Ca$_2$.
\item[\bf Fig. 3]     Weakly bound
  vibrational wavefunctions of the $(1)^1\Pi_g$ excited state for
    three different binding energies relevant for photoassociation 
    (a) and scattering wavefunction of two ground state calcium atoms
    (b).
\item[\bf Fig. 4]
    (a) The excited state population shows 
    Rabi oscillations  as the pulse energy is increased. The arrows
    indicate the pulsed energies which are employed for shaped pulses.
    (b) The projection of the final excited state wave packet onto the
    molecular levels, i.e. the excited state vibrational
    distribution after the pulse for a two-photon $\pi$-pulse and a
    two-photon $2\pi$-pulse. The arrows indicate the range of binding
    energies relevant in the discussion of the results for shaped
    pulses.
\item[\bf Fig. 5]
  Frequency-domain control with pulses shaped
    according to Eq.~(\ref{eq:phase}):
    The excited state population of atoms (filled circles) and of
    molecules (open symbols), normalized with respect to the results
    obtained for a transform-limited pulse, is shown as a function of
    the $\pi$-step position for (a) weak and (b) intermediate fields.
    Population transfer can be strongly suppressed, by $10^{-6}$, for
    the weak field (a). For the intermediate field (b), the dynamic Stark
    shift sets in yielding a maximum suppression on the order of
    $10^{-3}$.     
    When all molecular levels are considered, the last two levels (that
    have binding energies of less than 0.01$\,$cm$^{-1}$) dominate and no
    difference between atoms and molecules is observed. For 
    increasing binding energy of the molecular levels, and hence
    detuning from the atomic  resonance,  the two-photon transition
    becomes less dark.
\item[\bf Fig. 6]
  Time-domain control with pulses shaped according to
    Eq.~(\ref{eq:phasestrongfield}): 
    The excited state population of atoms (filled circles) and of
    molecules (open symbols), normalized with respect to the results
    obtained for a transform-limited pulse, is shown as a function of
    the duration
    of the corresponding transform-limited pulse for intermediate (a)
    and strong field (b). Two-photon transitions for atoms are
    more strongly suppressed than for  molecules. However, the
    suppression of the molecular transitions is too strong to achieve
    any significant photoassociation yield. For weakly bound levels no
    difference in the suppression with increasing binding energy is
    observed, i.e. the levels have very similar dynamic Stark shifts.
\end{list}

\clearpage

\end{document}